\documentclass[12pt]{article}

\pdfoutput=1
\usepackage{amssymb}
\usepackage{graphicx}
\usepackage{graphics}
\usepackage{amsmath}

\usepackage{bbold}

\def\d{\partial}

\newcommand{\be}{\begin{equation}}
\newcommand{\ee}{\end{equation}}
\newcommand{\ba}{\begin{align}}
\newcommand{\ea}{\end{align}}
\newcommand{\bg}{\begin{gather}}
\newcommand{\eg}{\end{gather}}
\newcommand{\bseq}{\begin{subequations}}
\newcommand{\eseq}{\end{subequations}}

\def\half{\frac{1}{2}}

\begin{document}
\title{On the minimal active-sterile neutrino mixing in seesaw
  type I mechanism with sterile neutrinos at GeV scale
}
\author{Dmitry Gorbunov$^{1,2}$, Alexander Panin$^{1}$\\
\mbox{}$^{1}${\small\em Institute for Nuclear Research of Russian Academy of
  Sciences, 117312 Moscow,
  Russia}\\  
\mbox{}$^{2}${\small\em Moscow Institute of Physics and Technology, 
141700 Dolgoprudny, Russia}  
}
\date{}

\maketitle

\begin{abstract} 
Renewed interest in GeV-scale sterile neutrinos capable of explaining
active neutrino oscillations via see-saw type I mechanism has been
expressed in several proposals of direct searches. Given this activity
we estimate the minimal values of sterile-active mixing angles 
provided one, two, or three sterile neutrinos are lighter than D-meson. 
\end{abstract}

\vskip 0.5cm
{\bf 1.} Neutrino oscillations definitely ask for some extension of
the Standard Model of particle physics (SM), and may be the simplest, 
yet complete and renormalizable, version is introducing 
three new Majorana massive
fermions, $N_I$, $I=1,2,3$, sterile with respect to all SM gauge
interactions. One can write down the most general renormalizable
Lagrangian 
\begin{equation}\label{2*}
{\cal L}=i{\bar N}_I \gamma^\mu \d_\mu N_I - \half\, 
M_I {\bar {N^c}}_I N_I - Y_{\alpha
  I} {\bar L}_\alpha {\tilde H} N_I + h.c.\;,
\end{equation}   
where $M_I$ are the Majorana masses and $Y_{\alpha I}$ stand for
Yukawa couplings with lepton doublets $L_\alpha$,
$\alpha=e,\,\mu,\,\tau$ and SM Higgs doublet $H$  
($\tilde H_a=\epsilon_{ab} H_b^*$,  $a=1,2$). When the Higgs field
gains vacuum expectation value $v=246$\,GeV, Yukawa couplings in
\eqref{2*} yield mixing between sterile $N_I$ and active $\nu_\alpha$
neutrino states. Diagonalization of the neutral fermion mass matrix
provides active neutrinos with masses $m_i$ and mixing which are
responsible for neutrino oscillation phenomena. 

With some hierarchy between model parameters $Y_{\alpha I}$ and
$M_I/v$, when the Dirac mass scale $Yv$ is well below the Majorana
mass scale $M$, active-sterile neutrino mixing angles are small,
$U \sim Yv/M\ll 1$, and active neutrino masses are double
suppressed, $m \sim U^2 M$. This is a type I seesaw mechanism
(for a review see e.g., Ref.\,\cite{Mohapatra:1998rq}) which explains
naturally why active neutrino mass scale is much lower than masses of
other SM particles. However, the Dirac mass scale and hence the size of
Yukawa couplings are not fixed from this reasoning: the seesaw
mechanism determines not $Y$ and $M$ but ratio $Y^2/M$. 

Then it is tempting to consider a variant of type I seesaw model,
where some of sterile neutrinos are at GeV scale. The variant can be
directly tested in particle physics experiments, where sterile
neutrinos appear in heavy hadron decays and subsequently decay in
light SM particles. Quite remarkably, in a part of parameter space
this model can explain not only neutrino oscillations but also baryon
asymmetry of the Universe via lepton number generation (leptogenesis)
in primordial plasma \cite{Drewes:2012ma}. In addition, it can even
provide with a dark matter candidate as light sterile neutrino of 1-50
keV mass; then, independently of dark matter production mechanism, the
successful leptogenesis requires two heavier sterile
neutrinos to be degenerate in mass. This pattern of seesaw type I is
known as $\nu$MSM (for neutrino minimal extension of the SM, see
details in e.g., \cite{Boyarsky:2009ix}). 

In the past century several dedicated searches for GeV-scale sterile neutrinos
were performed in fixed target experiments with negative results
\cite{Beringer:1900zz}. Recently, searches for the sterile neutrino signal
has been done by Belle Collaboration \cite{Liventsev:2013zz} 
in $e^+e^-$-collisions at KEK, similar studies are planned for LHCb
\cite{Bediaga:2012py} at CERN. Special investigation of sterile
neutrinos from kaon decays is undertaken in E494 experiment
\cite{Shaykhiev:2011zz} at BNL, and is suggested for T2K
experiment \cite{Asaka:2012bb}. GeV-scale sterile neutrinos are
considered in physical programs of proposed next generation long-base line
neutrino oscillation experiments: near detectors in LBNE\,\cite{Akiri:2011dv}, 
NuSOnG\,\cite{Adams:2008cm}, HiResM$\nu$\,\cite{Mishra:2008nx}.

Recently, an idea has been put forward \cite{Gninenko:2013tk} 
to construct a dedicated 
experiment and fully explore a part of $\nu$MSM parameter space, where
heavier sterile neutrinos responsible for leptogenesis are light
enough to emerge in $D$-meson decays. Later the idea has got a support  
and motivated a realistic proposal \cite{Bonivento:2013jag} 
of a beam-target experiment at CERN based on high-intensity 
SPS beam of 400\,GeV protons.   

Given the interest to the subject we estimate in this paper the minimal
values of mixings between sterile and active neutrinos, allowed by
type I seesaw mechanism, if some of sterile neutrinos are lighter than
2 GeV. One can argue that most sensitive to this model is a fixed-target
experiment, where produced by a proton beam hadrons decay into sterile
neutrinos. The obtained results are needed to estimate the
sensitivity of the future experiments required {\em to fully explore the
parameter space of type I seesaw models with sterile neutrinos} in
the interesting mass range.

\vskip 0.5cm 
{\bf 2.} 
It is convenient to adopt the bottom-up parametrization for the
$3\times 3$ Yukawa coupling matrix $Y_\nu$, 
firstly proposed in~\cite{Casas:2001sr},
\begin{equation}
\label{yukawas}
 Y_\nu \equiv \frac{i\sqrt{2}}{v}\,\sqrt{M_R}\,R\,\sqrt{m_\nu}\,
 U^\dag_{\text{PMNS}}\;,
\end{equation}
where $M_R \equiv \mathrm{diag}\{M_1,M_2,M_3\}$, 
$m_\nu \equiv \mathrm{diag}\{m_1,m_2,m_3\}$, $U_{\text{PMNS}}$ is the unitary 
Pontecorvo--Maki--Nakagawa--Sakata matrix and $R$ is a complex
orthogonal matrix, $R^T R = \mathbb{1}$. We take for
the active neutrino sector the central values of the combined 
fit \cite{Tortola:2012te} 
to neutrino oscillation data and all (still unknown) 
complex phases set to zero, 
\begin{subequations}
\begin{align}
&m_{atm} = 5.01 \times 10^{-2}~\text{eV}, \\
&m_{sol} = 8.73 \times 10^{-3}~\text{eV},\\
&\theta_{12} = 34.45^\circ\;, \\
&\theta_{23} = 51.53^\circ\;, \\
&\theta_{13} = 9.02^\circ\;,\\
\label{phases}
&\delta = \alpha_1 = \alpha_2 = 0\;.
\end{align}
\end{subequations}
Matrix $R$ can be parametrized as 
\begin{equation}
\label{matR}
R = \mathrm{diag}\{\pm 1, \pm 1, \pm 1\} \times
\begin{pmatrix}
c_2 c_1 & c_2 s_1 & s_2 \\
-c_3 s_1 - s_3 s_2 c_1 & c_3 c_1 - s_3 s_2 s_1 & s_3 c_2 \\
s_3 s_1 - c_3 s_2 c_1 & - s_3 c_1 - c_3 s_2 s_1 & c_3 c_2
\end{pmatrix}\;,
\end{equation}
where $c_i = \mathrm{cos}\,{z_i}$, $s_i = \mathrm{sin}\,{z_i}$ and
$z_i$ are three complex angles. Thus, the matrix of Yukawa couplings 
$Y_\nu$ depends on six dimensionless extra parameters, which do not
enter the active neutrino sector. Three other model parameters from
sterile neutrino sector are the three Majorana masses $M_I$. 
One can introduce the matrix of mixing angles between active and sterile
neutrinos by 
\begin{equation}
\label{angles}
U = \frac{v}{\sqrt{2}}\,M_R^{-1}\,Y_\nu\;.
\end{equation}

In a fixed-target experiment the sterile neutrinos are produced due to
mixing \eqref{angles} in weak decays of hadrons (if kinematically
allowed, $M_I<2$\,GeV). The same mixing is
responsible for sterile neutrino weak decays into SM particles, that
is the main signature accepted in sterile-neutrinos hunting.   
Therefore the number of signal events depends on the values of
$|U_{I\alpha}|^2$. For charmed hadrons as the main source of
sterile neutrinos, $|U_{I\tau}|^2$ may contribute to production
only (via decays of $\tau$-leptons from $D_s$-mesons), and is 
irrelevant for subsequent sterile neutrino decays due to kinematics. 
Thus, to be conservative, below we are interested in minimal values of
$|U_{Ie}|^2$ and $|U_{I\mu}|^2$, which determine the
maximal sensitivity of an experiment required to fully explore type I seesaw
model. 

{\bf 3.} 
We begin with a special situation, when two sterile neutrinos 
are degenerate in mass, $M_1 = M_2 \equiv {\cal M}$, 
while mass of the third sterile neutrino $M_3$ varies independently. 
If $M_3<2$\,GeV but ${\cal M}>2$\,GeV, the interesting values are
$|U_{3e}|^2$ and $|U_{3\mu}|^2$, however they both can take
zero values. It happens when $N_3$ mixes only 
with $\nu_\tau$ or does not mix at all with active neutrinos 
(then one of active neutrinos is massless). 

\begin{figure}[!htb]
\centerline{\includegraphics[width=0.5\textwidth]{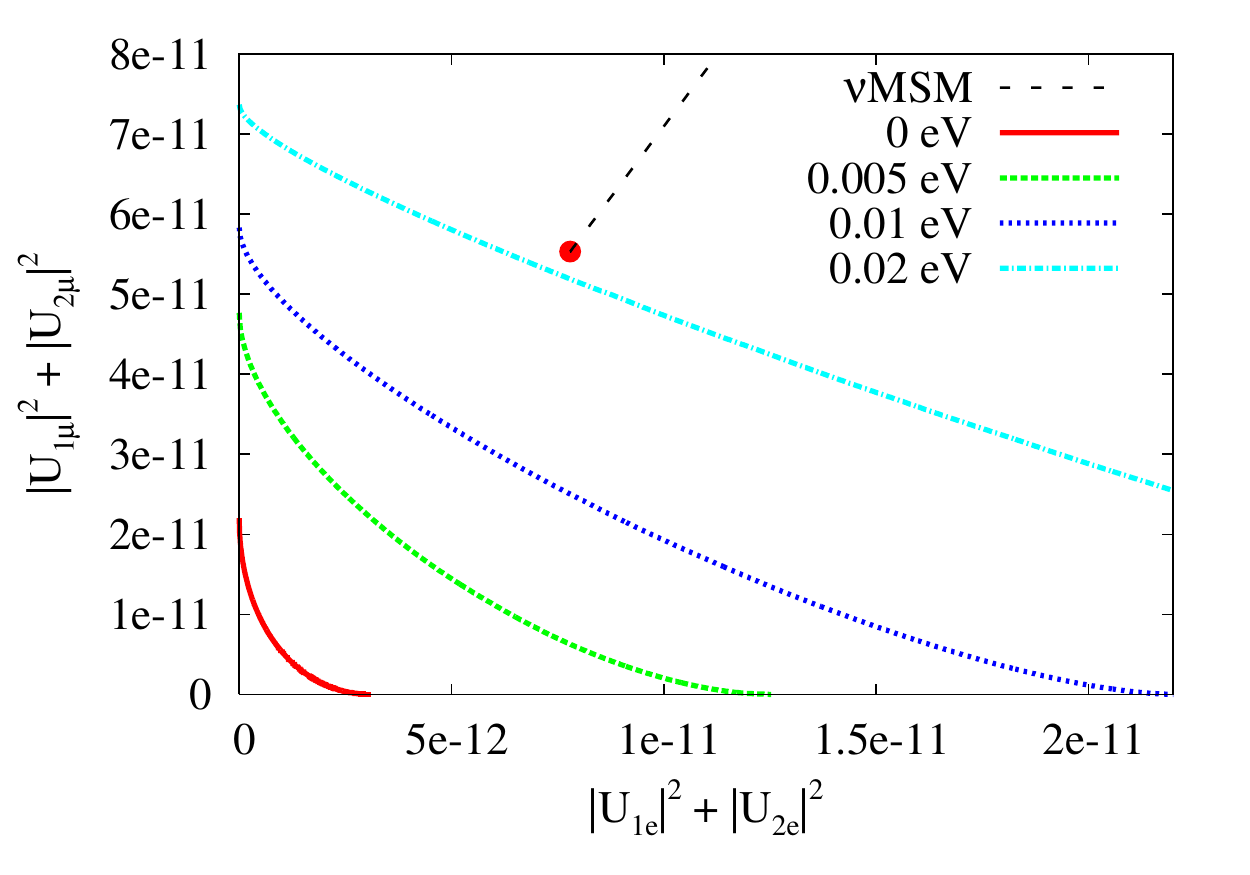} 
\includegraphics[width=0.5\textwidth]{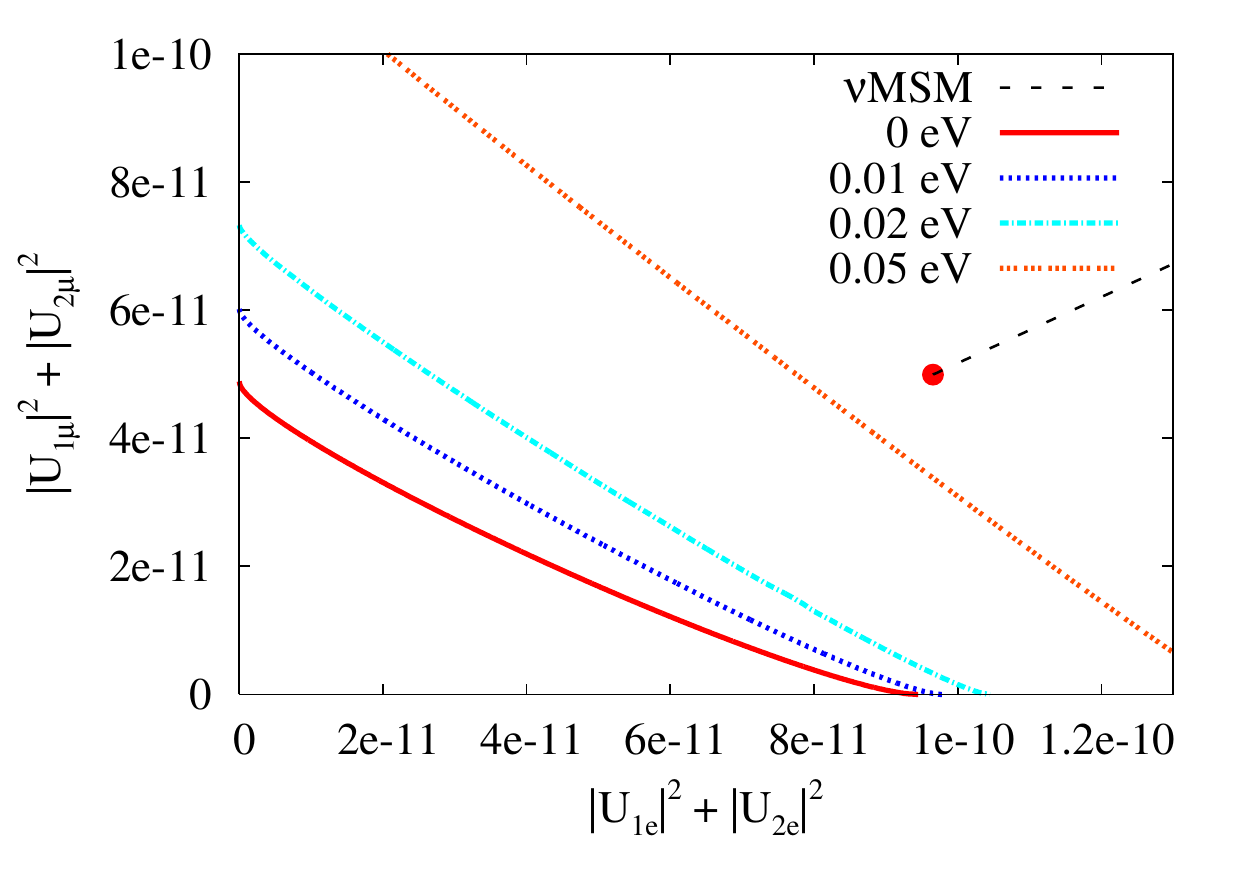}}
\centerline{\hspace{1cm}(a) \hspace{7.5cm} (b)}
\caption{\label{fig} The minimal values of mixings for
  $M_1=M_2=500$\,MeV and: (a)
normal hierarchy, (b) inverted hierarchy of active neutrino
masses. The different lines
correspond to different values of the lightest active neutrino 
mass $m_{\text{lightest}}$. On both plots the dashed line refers 
to the mixing in
$\nu$MSM \cite{Gorbunov:2007ak,Boyarsky:2009ix} as explained in the
main text.}
\end{figure}

In the opposite case, $M_3>2$\,GeV but ${\cal M}<2$\,GeV, 
our goal is to calculate minimal possible values of the sums 
\begin{subequations}
\label{sums}
\begin{align}
&{\cal U}_e = |U_{1e}|^2+|U_{2e}|^2\;, \\
&{\cal U}_\mu = |U_{1\mu}|^2+|U_{2\mu}|^2\;.
\end{align}
\end{subequations}
For ${\cal M}=500$\,MeV the numerical 
results are presented in Fig.\,\ref{fig}  
for two different hierarchies in active neutrinos masses. They are
obtained by scanning over parameters of matrix $R$ (see
Eq.\,\eqref{matR}) for a set of values of the lightest active neutrino
mass $m_{\text{lightest}}$. The values of ${\cal U}_{e,\,\mu}$ change 
with ${\cal M}$ as
${\cal U}_{e,\,\mu}\propto {\cal M}^{-1}$, and are independent of the
heaviest sterile neutrino mass $M_3$, in agreement with
Eqs.\,\eqref{yukawas},\,\eqref{angles}. Actually, these formulas imply
that the minimal values of ${\cal U}_{e,\,\mu}$ remain the same even for
the third sterile neutrino lighter than $2$\,GeV. This third neutrino
may either show up or be unobservable in the beam-target experiment (for
example, due to kinematics if $M_3$ is in keV region, or because when
both ${\cal M}<2$\,GeV and $M_3<2$\,GeV, $|U_{3e}|^2$ and
$|U_{3\mu}|^2$ can take zero values). With additional constraints
on the model parameters, minimal ${\cal U}_e$ and ${\cal U}_\mu$ generally
grow and certainly never drop below the lines presented in
Fig.\,\ref{fig}.  In particular, for the $\nu$MSM
\cite{Boyarsky:2009ix} (where the lightest sterile neutrino is dark
matter and almost decoupled from active neutrinos) the parameters are
so constrained that ${\cal U}_e$ and ${\cal U}_\mu$ 
for the two heavier
(almost) degenerate neutrinos \cite{Gorbunov:2007ak} are proportional
to each other as indicated in Fig.\,\ref{fig}. Minimal
(${\cal U}_e$, ${\cal U}_\mu$), marked by point in Fig.\,\ref{fig}, 
is well above the lower limit calculated for the unconstrained case
with $m_{\text{lightest}}=0$, relevant for $\nu$MSM .

Now we turn to more general case and split $N_1$ and $N_2$, so that
$M_1<M_2<2$\,GeV. We found numerically that lower limits on both
${\cal U}_e$ and ${\cal U}_\mu$ scale from the numbers given in
Fig.\,\ref{fig} as $\propto M_2^{-1}$. Note however, that when
neutrinos are of different masses, the signals are expected at
different masses (i.e. invariant masses of outcoming charged pion and 
lepton) and strengths in $e$,\,$\mu$ channels. They are shared by the two
sterile neutrinos in some proportions.  For example, it may happen,
that the signal in $e$-channel is saturated by the first sterile
neutrino, while the signal in $\mu$-channel is mostly due to the
second neutrino, and thus happens at different mass. The
generalization to the case of three neutrinos of similar masses at
(sub)GeV scale observable at beam-target experiment is
straightforward. 

\begin{figure}[!htb]
\centerline{\includegraphics[width=0.5\textwidth]{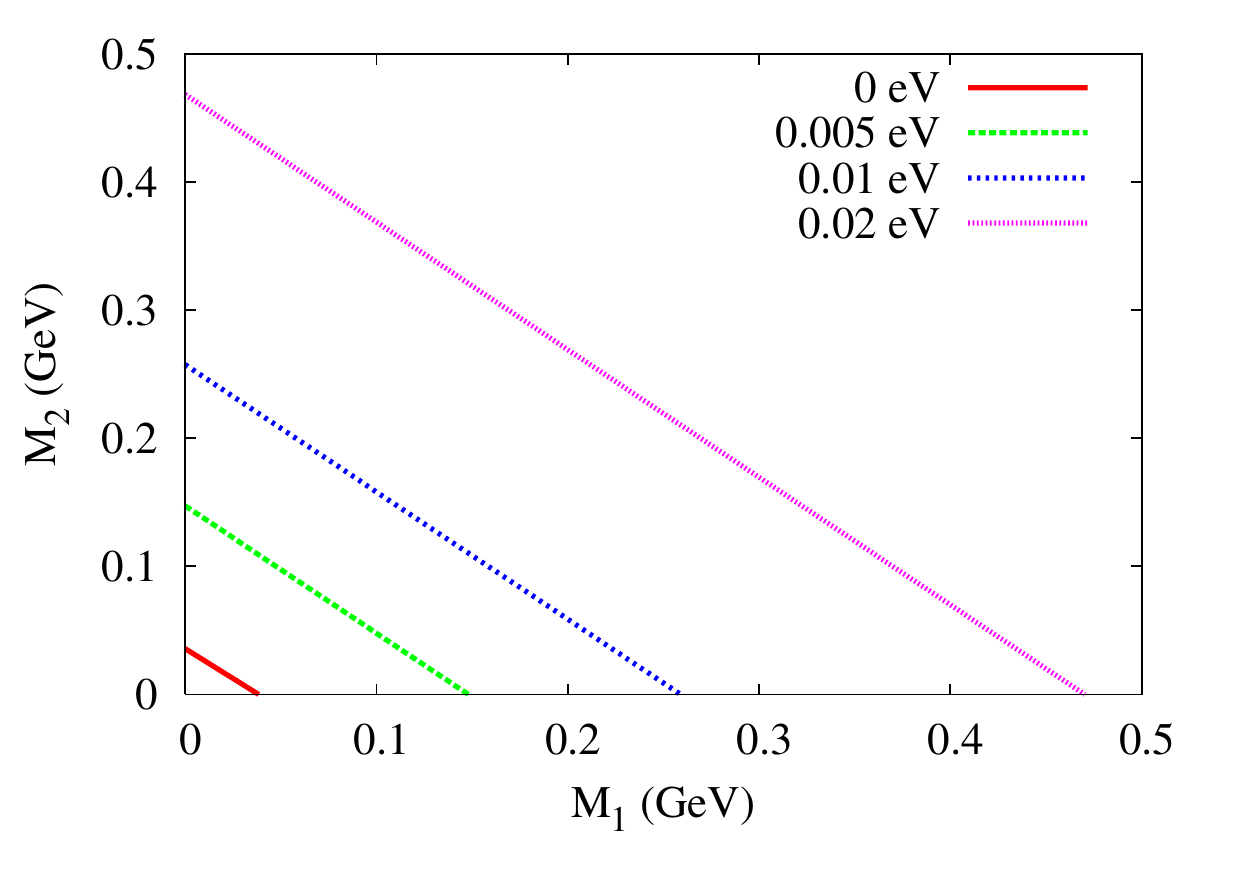} 
\includegraphics[width=0.5\textwidth]{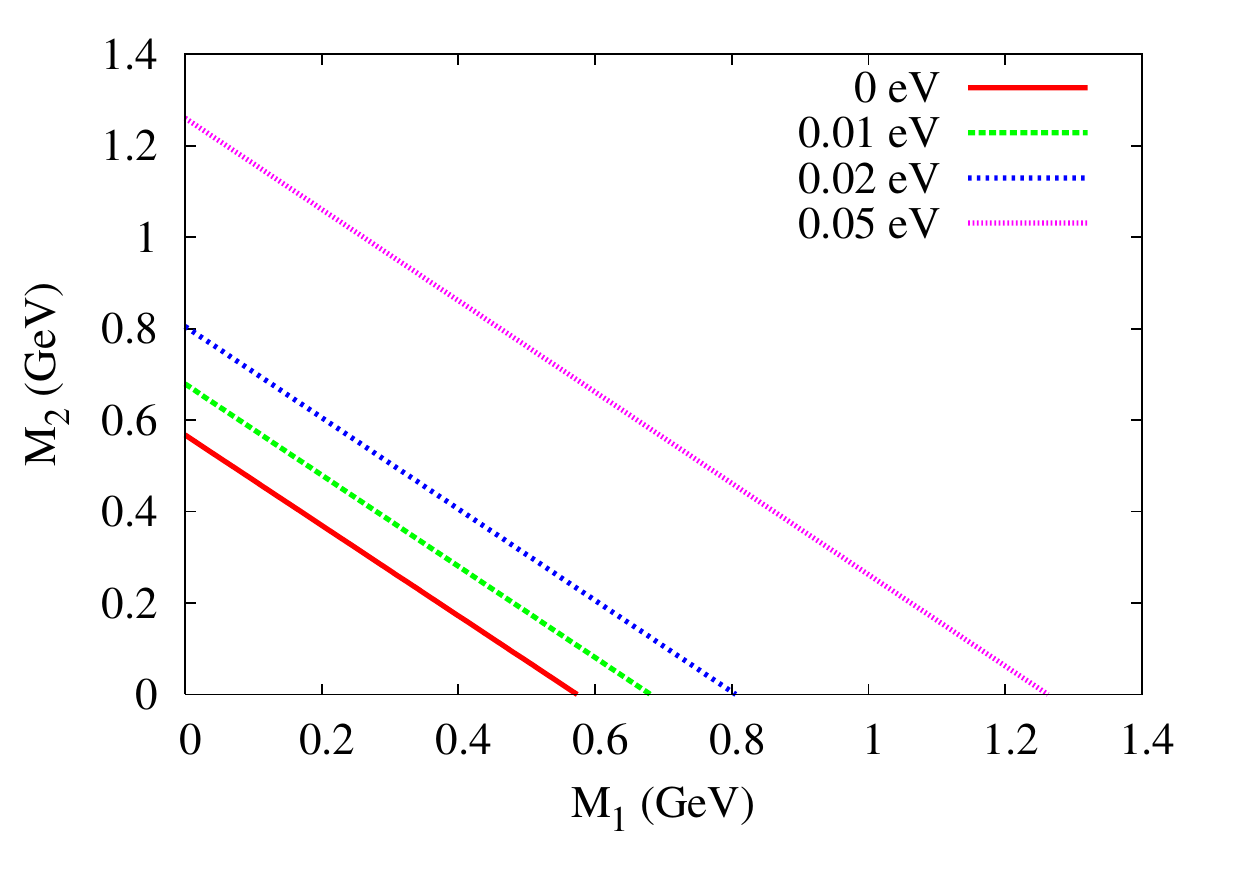}}
\centerline{\hspace{1cm}(a) \hspace{7.5cm} (b)}
\caption{\label{fig2} The regions below the lines will be excluded by
  an experiment achieving the sensitivity to mixing parameters of
  $U_c^2=2.5\times 10^{-11}$ for: (a)
normal hierarchy, (b) inverted hierarchy of active neutrino
masses. The different lines
correspond to different values of the lightest active neutrino 
mass $m_{\text{lightest}}$.}
\end{figure}

To illustrate the dependence on the sterile neutrino masses we outline
in Fig.\,\ref{fig2} 
the regions in $(M_1,M_2)$ space, where at least one of the squared
mixing values
$|U_{I\alpha}|^2$, $I=1,2$, $\alpha=e,\mu$, exceeds a given
value $U_c^2$. Reaching the sensitivity $U_c^2$, an
experiment rules out the seesaw model with sterile neutrino masses in
the corresponding region. 
The exclusion regions are separated by lines 
$M_1+M_2=f(m_{\text{lightest}})/U_c^2$.

Finally, we have checked that switching on complex
phases in active neutrino sector \eqref{phases} may change the obtained
estimates of {\em minimal mixing values} ${\cal U}_e$, ${\cal U}_\mu$
by some tens percent. This 
is due to suppression of one of mixings $|U_{I\alpha}|^2$, $I = 1,2$,
$\alpha = e, \mu$ occuring for particular sets of
angles and complex phases of $U_{\text{PMNS}}$ 
(see e.g. \cite{Ruchayskiy:2011aa}). This consideration completes the 
discussion. 

{\bf 4.} To conclude, we found minimal values of active-sterile
neutrino mixing for seesaw type I model for those sterile neutrinos
which are lighter than 2\,GeV. Present experimental upper limits on
the mixing can be found in Ref.\,\cite{Atre:2009rg}. 
The authors are indebted to F. Bezrukov, O. Ruchayskiy and M. Shaposhnikov
for useful discussions. 
The work is supported in part by the grant of the
President of the Russian Federation NS-2835.2014.2. 
The work of D.G. is supported in part by RFBR grant  
13-02-01127a.  
The work of A.P. is supported in 
part by the grant of the President of Russian
Federation MK-1754.2013.2.


\end{document}